\documentclass[aps,prb,twocolumn,superscriptaddress,amsmath,amssymb,amsfonts,graphics]{revtex4}

\pdfoutput=1
\usepackage{hyperref}
\hypersetup{pdfauthor={KSD Beach and FF Assaad},pdftitle={Coherence and metamagnetism in the two-dimensional Kondo lattice model}}

\usepackage{graphicx}
\usepackage{color}
\usepackage{bm}
\usepackage{bbold}
\newcommand{\svec}[1]{\bm{#1}}
\renewcommand{\vec}[1]{\mathbf{#1}}

\begin{document}

\title{Coherence and metamagnetism in the two-dimensional Kondo lattice model}
\author{K.\ S.\ D.\ Beach}
\affiliation{Institut\,\,f\"ur\,\,theoretische\,\,Physik\,\,und\,\,Astrophysik,\,\,Universit\"at\,\,W\"urzburg,\,\,Am Hubland,\,\,D-97074\,\,W\"urzburg,\,\,Germany}
\affiliation{Department\,\,of\,\,Physics,\,\,University\,\,of\,\,Alberta,\,\,Edmonton,\,\,Alberta\,\,T6G 2G7,\,\,Canada}
\author{F.\ F.\ Assaad}
\affiliation{Institut\,\,f\"ur\,\,theoretische\,\,Physik\,\,und\,\,Astrophysik,\,\,Universit\"at\,\,W\"urzburg,\,\,Am Hubland,\,\,D-97074\,\,W\"urzburg,\,\,Germany}

\begin{abstract}
We report the results of dynamical mean field calculations for the metallic Kondo lattice 
model subject to an applied magnetic field. High-quality spectral functions reveal that 
the picture of rigid, hybridized bands, Zeeman-shifted in proportion to the field strength, 
is qualitatively correct. We find evidence of a zero-temperature magnetization plateau,
whose onset coincides with the chemical potential entering the spin up hybridization gap.
The plateau appears at the field scale predicted by (static) large-$N$ mean field theory
and has a magnetization value consistent with that of $x=1-n_c$ spin-polarized heavy holes,
where $n_c < 1$ is the conduction band filling of the noninteracting system.
We argue that the emergence of the plateau at low temperature marks
the onset of quasiparticle coherence.
\end{abstract}


\maketitle

\section{Introduction}

The paramagnetic, metallic ground state of heavy fermion 
compounds~\cite{Stewart84,Lee86} can be interpreted as a Fermi liquid with 
low coherence temperature and large effective mass.~\cite{Auerbach86} 
The origin of the coherence temperature lies in the nature of the 
quasiparticles, which emerge as a coherent superposition of the Kondo screening 
clouds of  the individual magnetic impurities. Being a Fermi liquid, the heavy fermion
ground state is well captured by large-$N$ mean-field theories that predict
a renormalized band structure; at this level of approximation, the key 
properties---such as the effective mass  and the Fermi-surface topology---are reproduced.

Mean-field approaches that are static in time, however, do not properly incorporate 
Kondo screening and hence do not account for the many-body nature of the quasiparticles.
Such approaches become questionable when the system is subject to perturbations, 
such as large temperatures or strong magnetic fields, that have the potential
to destroy Kondo screening---hence, the  motivation to 
consider dynamical mean field theories (DMFT), where the Kondo effect is built in. 

The problem of a Kondo lattice insulator (symmetric conduction band at half-filling) 
in a magnetic field has previously been treated approximately using DMFT\cite{Ohashi05} 
and exactly using quantum Monte Carlo.~\cite{Beach04,Milat04} 
In that special case, the only important low-energy scale is the indirect gap separating upper 
and lower quasiparticle bands. In the metallic case (e.g., filling less than half), there is an 
additional, smaller energy scale given by the separation between the chemical potential 
and the top of the lower band.~\cite{Burdin00} This scale is generically small \emph{regardless of 
the conduction band filling} and controls the Fermi liquid properties. As we shall see,
it also sets the magnetic field scale for the onset of metamagnetic features.

The static mean field scenario for application of a magnetic field at zero 
temperature is as follows. As the field strength is increased, the spin up quasiparticle bands
descend. The spin up Fermi surface shrinks to a point and disappears as 
the lower band drops below the chemical potential.~\cite{Beach05,Daou06,Kusminskiy07} 
The resulting half-metallic state~\cite{Irkhin90}  has mean field parameters
that are locked at the values obtained before the disappearance of the
spin up Fermi surface and are completely insensitive to changes in the
field.~\cite{Beach05,Kusminskiy07}
As a result, the physical magnetization $-\partial \mathcal{F} / \partial B$ (where $\mathcal{F}$
is the free energy)
has a constant slope proportional to the difference between the Land\'{e} $g$-factors for the two species.
If the $c$- and $f$-electrons couple identically to the applied field, the system
is predicted to exhibit a magnetization plateau at a value that depends only 
on the conduction band filling $n_c$. Note that the dimensionless
magnetization $M = (n_{c,\uparrow}-n_{c,\downarrow})
+(n_{f,\uparrow}-n_{f,\downarrow})$ always shows the plateau behaviour, irrespective
of the values of $g_c$ and $g_f$. This is related to the fact that, in the locked state, 
the system behaves as a gas of $ x= 1-n_c$ fully spin-polarized, heavy-quasiparticle holes.  

Our DMFT results verify this scenario. We find that the magnetization profile begins to 
develop an inflection at temperatures on the order of the predicted coherence
temperature. This feature resembles a plateau smeared by temperature, and
we have verified that it sharpens as the temperature is lowered. Moreover, the 
apparent height of the plateau is consistent with the predicted magnetization value.
An analysis of the evolution of the spectral functions with applied field 
shows that the endpoints of the plateau do coincide with the chemical potential 
entering and leaving the hybridization gap. 

\section{ Model and methods}
We consider the Kondo lattice model (KLM) on a square lattice in an 
external magnetic field $B$:
\begin{equation} \label{EQ:KLM}
\begin{split}
   H   &= H_0 +  J \sum_{ \vec{i} }  \vec{S}^{c}_{\vec{i}} \cdot \vec{S}^{f}_{\vec{i}},\\
    H_0 &= \sum_{\vec{k},s} (\epsilon_{\vec{k}}-\mu)c^{\dagger}_{ \vec{k},s} c_{ \vec{k},s }
   - g \mu_B B \sum_{ \vec{i} } ( S_{\vec{i},z}^{c} + S_{\vec{i},z}^{f} ).
\end{split}
\end{equation}
Here, $c^{\dagger}_{\vec{k},s}$ creates a conduction electron in an extended orbital 
with wavevector $\vec{k}$ and spin projection $s={\uparrow,\downarrow}$ along an
axis of quantization chosen parallel to the applied field.
The tight-binding dispersion relation is $\epsilon_{\vec{k}} = -2t (\cos k_x + \cos k_y)$.
At each lattice site $\vec{i}$, a local spin-1/2 degree of freedom $\vec{S}^{f}_{\vec{i}}$ 
is coupled via $J$ to the $c$-electron spin density  $\vec{S}^{c}_{\vec{i}}=\tfrac{1}{2} 
\sum_{s,s'} c^{\dagger}_{ \vec{i}, s} \svec{\sigma}_{s,s'} c_{\vec{i}, s'} $
(represented with the aid of the Pauli spin matrices ${\svec{\sigma}}$).
An analogous expression can be written for $\vec{S}^{f}_{\vec{i}}$ using the 
localized orbital creation operators  $f^{\dagger}_{\vec{i},s}$.
Since the KLM forbids charge fluctuations on the $f$-orbitals, this representation
demands that we impose a strict constraint of one electron per $f$-orbital.

The DMFT approximation neglects spatial fluctuations and thereby omits the 
$\vec{k}$-dependence of the self-energy: i.e.,\ $\svec{\Sigma}_{s} (\vec{k}, i \omega_m) 
\rightarrow  \svec{\Sigma}_{s} (i \omega_m)$.~\cite{Georges96,Maier05}
Being local, the  self-energy  is that of a single impurity in a bath described by 
the free Hamilton operator $\mathcal{H}_0 
=  \sum_{ \vec{k},s } \tilde{\epsilon}_{\vec{k},s}
   c^{\dagger}_{ \vec{k}, s} c_{ \vec{k}, s } 
   - \tilde{g}_f \mu_B B S_{i,z}^{f}$. The corresponding local Green bath function is given by 
\begin{equation}
\svec{\mathcal{G}}_{s}(i \omega_m)  = 
\begin{pmatrix}
	\mathcal{G}^{cc}_{s} (i \omega_m) & 0 \\
	0  & \mathcal{G}^{ff}_{s} ( i \omega_m ) 
\end{pmatrix}
\end{equation}
using $\mathcal{G}^{cc}_{s} (i \omega_m) = 
-\int_{0}^{\beta} {\rm d} \tau e^{i \omega_m \tau} \langle 
c^{\phantom{\dagger}}_{\vec{0},s}(\tau) 
c^{\dagger}_{\vec{0},s} \rangle_{\mathcal{H}_0}$ and the
equivalent definition for $\mathcal{G}^{ff}$.
The prerequisite for the implementation of DMFT is the ability to solve, for a given bath, the 
Kondo model,
\begin{equation}
\mathcal{H}   =  \mathcal{H}_0   + J   \vec{S}^{c}_{\vec{0}} \cdot \vec{S}^{f}_{\vec{0}}.
\end{equation}
In our calculations, we have opted for the Hirsch-Fye approach. 
Following Ref.~\onlinecite{Capponi00},
we expand the Hilbert space to permit charge fluctuations on the $f$-sites and 
replace the Kondo term by
\begin{multline}
\label{EQ:KLM_HS}
    -\frac{J}{4}
        \biggl( \sum_{s} c^{\dagger}_{\vec{0}, s} 
         f^{\phantom{\dagger}}_{\vec{0}, s} + \text{h.c.} \biggr)^2\\
        +U \bigl( f^{\dagger}_{\vec{0}, \uparrow} f^{\phantom{\dagger}}_{\vec{0}, \uparrow}  - 1/2 \bigr)
           \bigl( f^{\dagger}_{\vec{0}, \downarrow} f^{\phantom{\dagger}}_{\vec{0}, \downarrow}  - 1/2 \bigr).
\end{multline}
At the expense of two discrete Hubbard-Stratonovich fields, we can decompose the (perfect 
square) hybridization and Hubbard terms, and readily implement the above interaction within the 
framework of the Hirsch-Fye algorithm.~\cite{Capponi00}
In the limit $U \rightarrow \infty $, charge fluctuations  on the $f$-orbitals are  suppressed 
and the squared hybridization reduces to the desired Kondo term. The efficiency  of 
this approach lies in the fact that
    $U ( f^{\dagger}_{\vec{0}, \uparrow} f_{\vec{0}, \uparrow}  - 1/2 )
       ( f^{\dagger}_{\vec{0}, \downarrow} f_{\vec{0}, \downarrow}  - 1/2 )$ is a conserved 
quantity~\cite{Capponi00} for the considered class of $\mathcal{H}_0$, which does not include 
hybridization terms between the bath and impurity orbital. Typically, $\beta U  \simeq 30$ 
suffices to ensure that  double occupancy drops down to
$ \langle  f^{\dagger}_{\vec{0}, \uparrow} f_{\vec{0}, \uparrow}
f^{\dagger}_{\vec{0}, \downarrow} f_{\vec{0}, \downarrow}  \rangle =
 0.0005  \pm 0.0005 $ and that
$ \langle  \sum_{s}  f^{\dagger}_{\vec{0}, s} f_{\vec{0}, s}  \rangle = 1 $ to 
the same precision.
Since  charge fluctuations on the $f$-sites are suppressed by the Hubbard interaction,  the local 
Green function, as obtained from the Hirsch-Fye alogorithm,  is diagonal in the large-$U$ limit 
and is given by
\begin{equation}
	\vec{G}_{s}( i \omega_m ) = \frac{1} {\svec{\mathcal{G}}_{s}^{-1}(i \omega_m) - 
   \svec{\Sigma}_{s}(i \omega_m) }.
\end{equation}

Self-consistency requires that the local Green 
function,  as  determined from the effective impurity problem,  matches that of the lattice: 
\begin{eqnarray}
\label{EQ:DMFT}
        \vec{G}_{s} (i \omega_m) & \equiv & 
\frac{1}{N} \sum_{\vec{k}} \frac{1} { \vec{G}_{0,s}^{-1} (\vec{k},i \omega_m) - 
         \svec{\Sigma}_{s} (i \omega_m) }
\nonumber \\
 & = & \frac{1}{
      \svec{\mathcal{G}}_{s}(i \omega_m) - \svec{\Sigma}_{s}(i \omega_m) }.
\end{eqnarray}
Here, $\vec{G}_{0,s}(\vec{k},i\omega_m) $ is the free lattice Green function 
[corresponding to $H_0$ in Eq.~\eqref{EQ:KLM}].~\footnote{Relaxing the constraint and using 
Eq.~\eqref{EQ:KLM_HS} to emulate the Kondo interaction allows a weak coupling perturbative 
derivation of the self-consistency. One can then take the limit $U \rightarrow \infty$ to 
recover the Kondo model.}

Hence, for a given bath Green function $\svec{\mathcal{G}}_{s} (i \omega_m)$,
we extract the self-energy from the impurity solver and, exploiting Eq.~\eqref{EQ:DMFT},
recompute the  bath Green function with
\begin{multline}
     \svec{\mathcal{G}}_{s}^{-1}(i \omega_m) =
       \svec{\Sigma}_{s} (i \omega_m)   \\
 +\biggl( \frac{1}{N} \sum_{\vec{k}} \frac{1}{\vec{G}_{0,s}^{-1} 
              (\vec{k},i \omega_m) - 
         \svec{\Sigma}_{s} (i \omega_m) } \biggr)^{-1}.
\end{multline}
The procedure is repeated until convergence is achieved.
As noted above, in the limit $ U \rightarrow 
\infty $ the local Green function, self-energy, and bath Green function are all diagonal.
Accordingly, the self-consistency [see Eq.~\eqref{EQ:DMFT}] is equivalent to
two scalar equations:
% \begin{multline}
% \frac{1}{N}   \sum_{k} \frac{1}  { G_{0,\sigma}^{cc,-1} (\vec{k},i \omega_m) - 
%           \Sigma_{\sigma}^{cc}  (i \omega_m) }\\
%   =  \frac{1}
%       {  \mathcal{G}_{0,\sigma}^{cc,-1} (i \omega_m) - \Sigma^{cc,\sigma}(i \omega_m) },
% \end{multline}
\begin{gather}
\frac{1}{N}   \sum_{k} \frac{ {\mathcal{G}_{0,s}^{cc}}^{-1} (i \omega_m) - \Sigma^{cc}_{s}(i \omega_m)}  { {G_{0,s}^{cc}}^{-1} (\vec{k},i \omega_m) - 
          \Sigma_{s}^{cc}  (i \omega_m) }
  =  1,\\
  G_{0,s}^{ff} (i \omega_m) = \mathcal{G}^{ff}_{0,s}(i \omega_m).
\end{gather}
The last equality follows from the fact that $ G_{0,s}^{ff}(\vec{k},i \omega_m) $ 
has no $\vec{k}$ dependence.  

Having determined the self-energy, one can readily compute the single particle Green function
$G^{cc}_{s}(\vec{k},i \omega_m )$ and extract the spectral function $A(\vec{k},\omega)$ with a 
stochastic analytical continuation method.~\cite{Sandvik98,Beach04a}

\begin{figure}
\begin{center}
\includegraphics{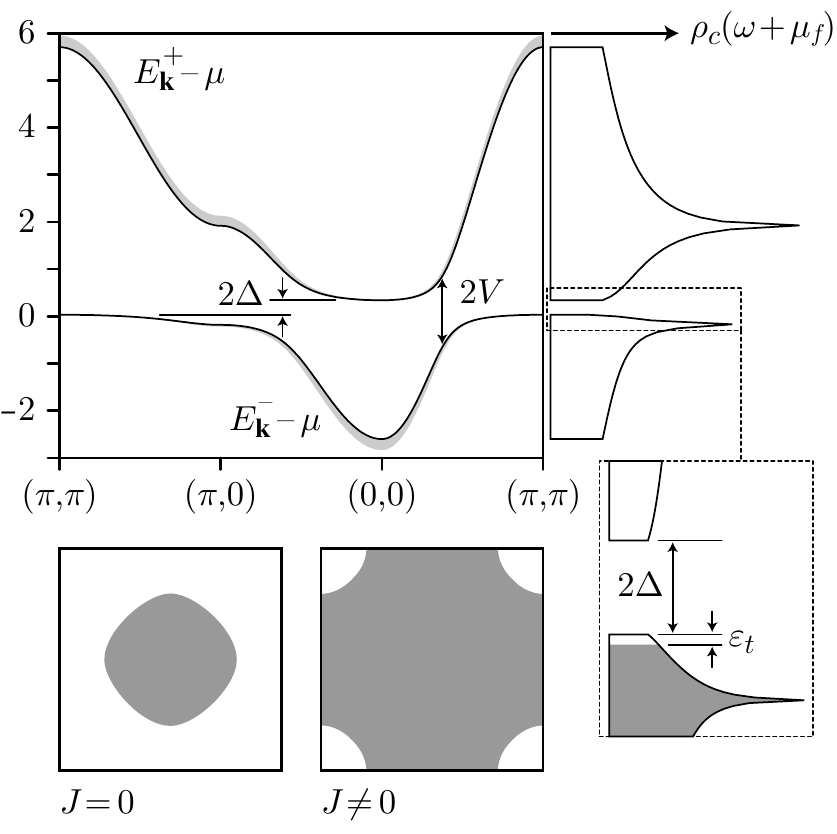} 
\end{center}
\caption{\label{FIG:bands}The upper and lower hybridized bands $E_{\vec{k}}^\pm$ are plotted over a trajectory
bounding one octant of the Brillouin zone. The grey line-width is proportional to the  
$c$-electron spectral weight. 
The indirect gap $2\Delta$ and the optical gap $2V$
are indicated. The corresponding $c$-electron density of states
$\rho_c(\omega)$ is plotted to the right on the same energy scale.
In magnification, the region around the gap is illustrated with the
filled Fermi sea shaded in grey, revealing a small amount of headroom
$\varepsilon_t \ll 2\Delta$ in the lower band. In the lower two panels,
the small, noninteracting Fermi surface is contrasted with the large
interacting one.
}
\end{figure}

\begin{figure}
\begin{center}
\includegraphics{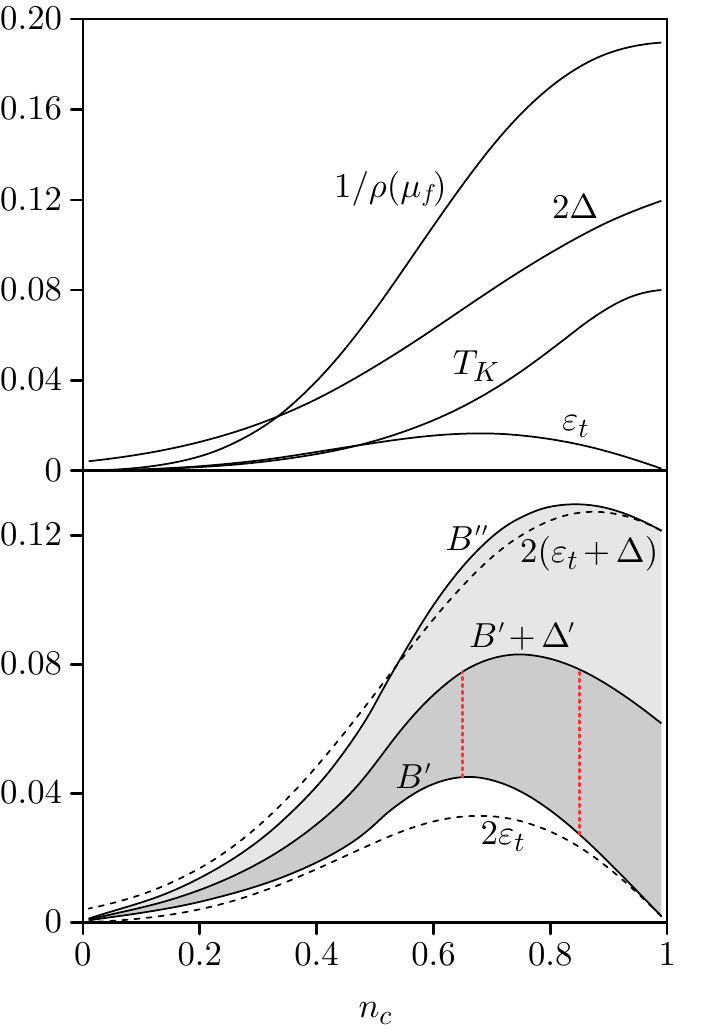} 
\end{center}
\caption{\label{FIG:scales}
(Upper panel) As a function of the bare filling $n_c$, the inverse quasiparticle 
DOS $1/\rho(\mu_f)$ and the indirect gap $2\Delta$ are compared to the energy scales 
defined by the Kondo temperature $T_{\text{K}}$ (with $k_B=1$) and the 
width $\varepsilon_t$ of unfilled states at the 
top of the lower band. (Lower panel) The dashed lines at $2\epsilon_t$ and $2(\epsilon_t + \Delta)$
mark the plateau endpoints that arise from a purely rigid shifting of the bands (in units
where $\mu_B g =1$). 
The shaded areas between $B'$ and $B''$ indicate the extent of the locked region
as calculated using the fully self-consistent mean field equations. The discrepancy
is a result of higher order changes in the mean field parameters. We expect that
the inclusion of $V$ phase fluctuations restricts plateau formation 
to between $B'$ and $B'+\Delta'$ (roughly)
and leads to a slow disintegration of the heavy fermion state between $B'+\Delta'$ and $B''$.
The vertical dashed (red) lines between $B'$ and $B'+\Delta'$ at $n_c = 0.65$ and $n_c = 0.85$
appear again as horizontal dashed (red) lines in Fig.~\ref{FIG:plateaux}.
}
\end{figure}

\begin{figure}
\begin{center}
\includegraphics{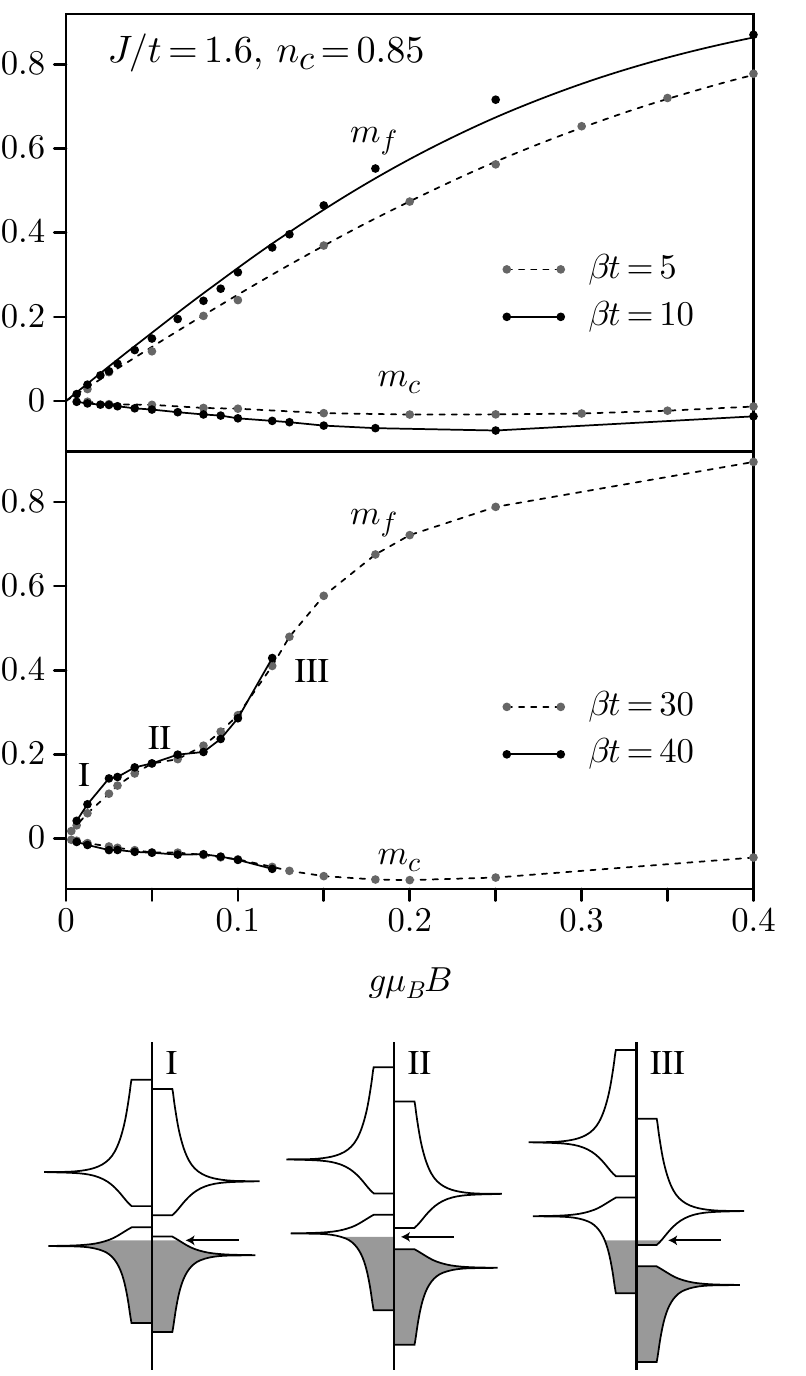} 
\end{center}
\caption{\label{FIG:magnetization} 
The magnetization of conduction $m_c$ and local moments $m_f$,
computed via DMFT, are plotted as a function  of magnetic field at various 
temperatures. In the upper panel, the lines running through
the $f$-electron data points are fit to the forms 
$\tanh(S^{\text{eff}}  g \mu_B B \beta)$. At $\beta t = 5$, $2S^{\text{eff}} = 1.03$ and 
at  $\beta t = 10$, $2S^{\text{eff}}  = 0.65$. In the lower panel, the magnetization
profile is marked with three regions. A sketch of the spin resolved density of states as deduced 
from the  single particle spectral function  (See Fig.~\ref{FIG:Aom_B}) is given in each region. 
 }
\end{figure}

\section{Mean field picture}
\label{SEC:mean-field}
Free electrons moving via nearest-neighbour hopping on the square lattice
have a density of states (DOS)
\begin{equation} \label{EQ:bareDOS}
\rho_0(\omega) = \frac{1}{2\pi^2t}\mathcal{K}\Bigl(\!\sqrt{1-(\omega/4t)^2}\,\Bigr)\theta(16t^2-\omega^2),
\end{equation}
written in closed form in terms of an elliptic integral of the first kind, $\mathcal{K}(x)$.
The function has support in the region $-4t < \omega < 4t$ and thus a bandwidth $W=8t$.
It is smooth and continous, except at the step-like band edges and at $\omega =0$
(a van Hove singularity), where it diverges as
$\rho_0(\omega) \sim \frac{1}{2\pi^2t}\log\frac{16t}{|\omega|}$.
[Numerical evaluation of Eq.~\eqref{EQ:bareDOS} can be carried out very efficiently
using series expansion; see Appendix~\ref{SEC:DOS}.]

At the mean-field level, the interacting Kondo lattice system is modelled by 
a bilinear effective Hamiltonian that includes both $c$- and $f$-electrons.
This can be obtained from the saddle-point of the Hubbard-Stratonovich 
decomposition of Eq.~\eqref{EQ:KLM_HS}:
\begin{equation}
\begin{pmatrix}
c^\dagger_{\vec{k}} & f^\dagger_{\vec{k}}
\end{pmatrix}\!
\begin{pmatrix}
\epsilon_{\vec{k}}\!-\!\mu_c\!-\!\tfrac{1}{2}g\mu_BB\sigma_z & -V\\
-V & -\mu_f\!-\!\tfrac{1}{2}g\mu_BB\sigma_z 
\end{pmatrix}\!
\begin{pmatrix}
c_{\vec{k}} \\ f_{\vec{k}}
\end{pmatrix}\!.
\end{equation}
In the usual way, hybridization between the $c$-electrons and the dispersionless 
band of $f$-electrons leads to a renormalized, quasiparticle dispersion
\begin{equation} \label{EQ:qp-dispersion}
E^{\pm}_{\vec{k},s} = \frac{1}{2}\biggl[
\epsilon_{\vec{k}} - sg\mu_BB \pm \sqrt{(\epsilon_{\vec{k}}-b)^2+4V^2}
\biggr]
\end{equation}
and a quasiparticle DOS 
$\rho(\omega) = \rho_c(\omega) + \rho_f(\omega)$,
where
\begin{equation} \label{EQ:qp-dos}
\rho_c(\omega) = \rho_0(\omega-V^2/\omega+b)
%\sum_{i=1}^4(-1)^i\theta(\omega-\omega_i)
\end{equation}
and $\rho_f(\omega) = (V^2/\omega^2)\rho_c(\omega)$. Here $b = \mu_c-\mu_f$ is the chemical energy to
transmute the $c$- and $f$-electron character of a particle, and $V$ is a hybridization energy
determined self-consistently via $V \sim J\sum_s\langle c^\dagger_{\vec{k},s} f_{\vec{k},s}\rangle$.
The spectral weight vanishes outside the lower $[\omega_1,\omega_2]$ and 
upper $[\omega_3,\omega_4]$ hybridized bands, where 
$\omega_i$ ($i=1,2,3,4$) denote the four ordered roots of 
$\pm 4t = \omega-V^2/\omega + b$.

The mean field equations can be written compactly as
\begin{equation} \label{EQ:MeanField}
\sum_s \int_{I}\!d\omega\,\rho_c(\omega)f(\omega-\mu_{f,s})
\begin{Bmatrix} 1 \\ V^2/\omega^2 \\ -\mathcal{J}/2\omega
\end{Bmatrix}= \begin{Bmatrix} n_c \\ 1 \\ 1 \end{Bmatrix},
\end{equation}
where the integral is taken over the disjoint interval 
$I = [\omega_1,\omega_2]\cup[\omega_3,\omega_4]$,
and $\mu_{f,s}$ is shorthand for $\mu_f \pm g\mu_BB/2$.
The first two equations fix the $c$- and $f$-electron occupation (to $n_c$ and 1, respectively),
and the third enforces the self-consistency condition on $V$. We have written $\mathcal{J}
= (\text{const.})\times J$ to allow for the fact that different mean field decompositions
lead to a different numerical prefactor.
In each spin channel, $2\Delta = \omega_3-\omega_2$ is the smallest indirect gap,
and $2V$ is the threshold for optical excitations.~\cite{Dordevic01,Millis87}
Figure~\ref{FIG:bands} illustrates the zero-temperature, zero-magnetic-field solution 
appropriate for band filling $0 < n_c < 1$. An additional energy scale
$\varepsilon_t = \omega_2 - \mu_f$, representing the headroom at the
top of the lower band, is indicated.

An artifact of the mean field treatment is that the
hybridization matrix element has an anomolous expectation value
that vanishes with heating at a second-order phase transition. 
(The true Kondo physics is that of a crossover.)
There is a critical temperature $T_{\text{c}}$ such that
as $T \to T_{\text{c}}$ from below, 
 $V,\mu_f \to 0$ and $b\to\mu_c$. As the quasiparticles
 disintegrate into their separate $c$- and $f$-character
 consituents, the corresponding densities
 of states return to their free values:  
 $\rho_c(\omega) \to \rho_0(\omega+\mu_c)$
and $\rho_f(\omega) \to \delta(\omega)$.
In this limit, Eq.~\eqref{EQ:MeanField} reduces to
%\begin{multline}
%\frac{2}{\mathcal{J}} = \int_{-4t-\mu_c}^{4t-\mu_c}\!d\omega\,\biggl[
%\rho_0(\omega+\mu_c)\frac{\tanh(\omega/2T_{\text{c}})}{\omega}\\
%-\frac{\rho_0(\omega+\mu_c)-\rho_0(\mu_c)}{\omega}\biggr]-\rho_0(\mu_c)\log\frac{4t-\mu_c}{4t+\mu_c}
%\end{multline}
\begin{multline}
\frac{2}{\mathcal{J}} = \int_{-4t}^{4t}\!d\omega\,\biggl[
\rho_0(\omega)\frac{\tanh[(\omega-\mu_c)/2T_{\text{c}}]}{\omega-\mu_c}\\
-\frac{\rho_0(\omega)-\rho_0(\mu_c)}{\omega-\mu_c}\biggr]-\rho_0(\mu_c)\log\frac{4t-\mu_c}{4t+\mu_c},
\end{multline}
where $\mu_c(n_c)$ takes its noninteracting value, determined implicitly by
\begin{equation}
2\int_{-4t}^{4t}\rho_0(\omega)f(\omega-\mu_c) = n_c.
\end{equation}

The critical temperature scales as $T_{\text{c}} \sim \alpha W$,
where the constant of proportionality is a function of the band filling alone.
(For a flat DOS, $T_{\text{c}} = 0.567n_c\times\alpha W$.) The small parameter
$\alpha = e^{-1/\mathcal{J}\rho_0(\mu_c)}$ renormalizes the bandwidth down 
to the Kondo scale; hence it is natural to identify a Kondo temperature 
$T_{\text{K}} \equiv T_{\text{c}}$.
On the square lattice, the value of $T_{\text{K}}$ is always nonzero for $n_c > 0$.
$T_{\text{K}}$ is compared to the important zero-temperature energy
scales in the upper panel of Fig.~\ref{FIG:scales}. A value
$\mathcal{J}/t=1.631$ was chosen
so that $\partial M/ \partial B\bigr\rvert_{B=0} = \rho(\mu_f)$
matches the DMFT results for $J/t=1.6$
and $n_c=0.85$. (We set $g\mu_B = 1$ for the remainder of
this section.)

At zero temperature, where the hybridization is at its strongest,
the Fermi function in Eq.~\eqref{EQ:MeanField} cuts off the 
integration at $\mu_{f,s}$. There are
three possibilities: (I) $\omega_1 < \mu_{f,s} < \omega_2$;
(II) $\omega_1 < \mu_{f,\downarrow} < \omega_2$ and
$\omega_2 < \mu_{f,\uparrow} < \omega_3$;
(III) $\omega_1 < \mu_{f,\downarrow} < \omega_2$ and
$\omega_3 < \mu_{f,\uparrow} < \omega_4$.
In cases I and III, both spin up and spin down quasiparticles
have a Fermi surface. In case II, the chemical potential lies
in the spin up hybridization gap. As a consequence, the upper limit of integration
in the spin up channel is $\omega_2$, and $\mu_{f,\uparrow}$ no longer enters the equations.
The magnetic field enters only indirectly through $\mu_{f,\downarrow}$,
and thus the system becomes insensitive to changes in magnetic field.
In particular, this leads to a plateau in the magnetization.
Note that $\mu_{f,\downarrow} = \text{const.}$ and hence $\mu_{f,\uparrow} = \text{const.} + B$,
which means that the upper band descends at twice the rate it does in
case I, where
$\mu_{f,\uparrow} \sim \mu_{f}\rvert_{B=0} + B/2$.
The plateau terminates when $\mu_{f,\uparrow}$ reaches the
lower edge of the upper hydridized band.

The leading edge of the plateau can be determined by solving for the field $B'$
at which $\mu_{f,\uparrow} = \omega_2$ and $\mu_{f,\downarrow} = \omega_2-B'$.
Since the mean field parameters are locked beyond $B'$, the far edge is simply
$B'' = B' + 2\Delta'$, where $2\Delta'$ is the size of the indirect hybridization gap
at $B=B'$ (and its size throughout $B' < B < B''$). 
This argument for the location of the far edge is flawed in one respect:
when $B > B''$, the system is expected to enter region III, but
the mean field equations in that region turn out to be 
pathological ($V$ begins to increase rapidly with $B$). Indeed,
energetic considerations tell us that for $B = B''[1 - O(\alpha^2)]$
the $V=0$ normal state becomes energetically favourable.~\cite{Beach05} 
Nonetheless, this apparent first-order collapse to the normal state cannot survive the
inclusion of phase fluctuations in $V$, which smooths out the destruction
of the heavy fermion state over a width $\Delta$. Hence, we expect
to find a plateau in $B' < B < B' + \Delta'$ and a crossover to the
normal state in $B' + \Delta' < B < B'' = B' + 2\Delta'$.
The predicted extent of the plateau is shown in the lower panel of Fig.~\ref{FIG:scales}. 

The mean field picture describes $x=1-n_c$ heavy holes, which spin-align with
the applied field and become completely polarized at $B'$. 
Along the plateau, the value of the magnetization is
$M = (n_{c\uparrow} + n_{f\uparrow}) - (n_{c\uparrow} + n_{f\uparrow}) = x[1-O(\alpha^2)]$.
This follows because
$n_{c\uparrow} \simeq n_{f\uparrow} \simeq 1$ and
$n_{c\downarrow} \simeq n_{f\downarrow} \simeq 1-x/2$ when the chemical
potential sits in the spin up hybridization gap.
In the Kondo limit ($\mathcal{J} \ll W$), $\alpha$ is exponentially small
and corrections of order $\alpha^2$ are completely negligible. Hence,
the plateau has a height of almost exactly $M=x$.

\section{ DMFT results }

The mean-field picture of the metamagnetic transition and the associated  change in the 
Fermi surface topology is a direct consequence of coherence.  
To confirm this, we have carried out temperature and magnetic field DMFT scans at $J/t = 1.6$ and 
electron density $n_c = 0.85$ (see Fig.~\ref{FIG:magnetization}). 
This choice of parameters sets the  single impurity Kondo 
scale to $ T_K/t \simeq 0.09$.\footnote{We have extracted the single impurity Kondo temperature 
by carrying out a data collapse of the impurity spin susceptibility}
At temperature scales larger than the Kondo scale, one 
expects the local moments to be essentially free with magnetization 
\begin{equation}
\label{EQ:free_moment}
       m_f \equiv 2 \langle S^f_{\vec{i},z} \rangle = 
       \tanh\bigl( S^{\text{eff}} \beta g \mu_B B\bigr),
\end{equation} 
where $S^{\text{eff}}$ allows for a renormalization of the impurity moment.
At $ \beta t  = 5$,  the  DMFT data  (see top panel of Fig.~\ref{FIG:magnetization}) 
for $m_f$ follows  this form with $S^{\text{eff}} \simeq 1/2$
to considerable accuracy, thereby showing that the system is in the 
high temperature  local moment regime.     
The polarization of the  conduction electrons is opposed to the 
applied magnetic field because the antiferromagnetic  Kondo interaction generates 
a negative effective magnetic field.~\cite{Beach04, Kusminskiy07}
In the vicinity of the Kondo temperature, at $ \beta t = 10$, the  magnetization curve 
$m_f$ can roughly be accounted  for with the free moment form  of 
Eq.~\eqref{EQ:free_moment}, albeit  with $S^{\text{eff}} \simeq 1/3$.  This 
reduction is consistent with the onset of Kondo screening. 
At temperatures $\beta t  > 30 $,  which we argue are well below the coherence 
temperature, three distinct regions denoted by I, II, and III are apparent in the 
data, as shown in Fig.~\ref{FIG:magnetization}.
Our understanding of  those features relies on the single particle spectral function  below the 
coherence temperature and the associated change in the Fermi surface topology as a function 
of the magnetic field.  

\begin{figure}
\begin{center}
\includegraphics{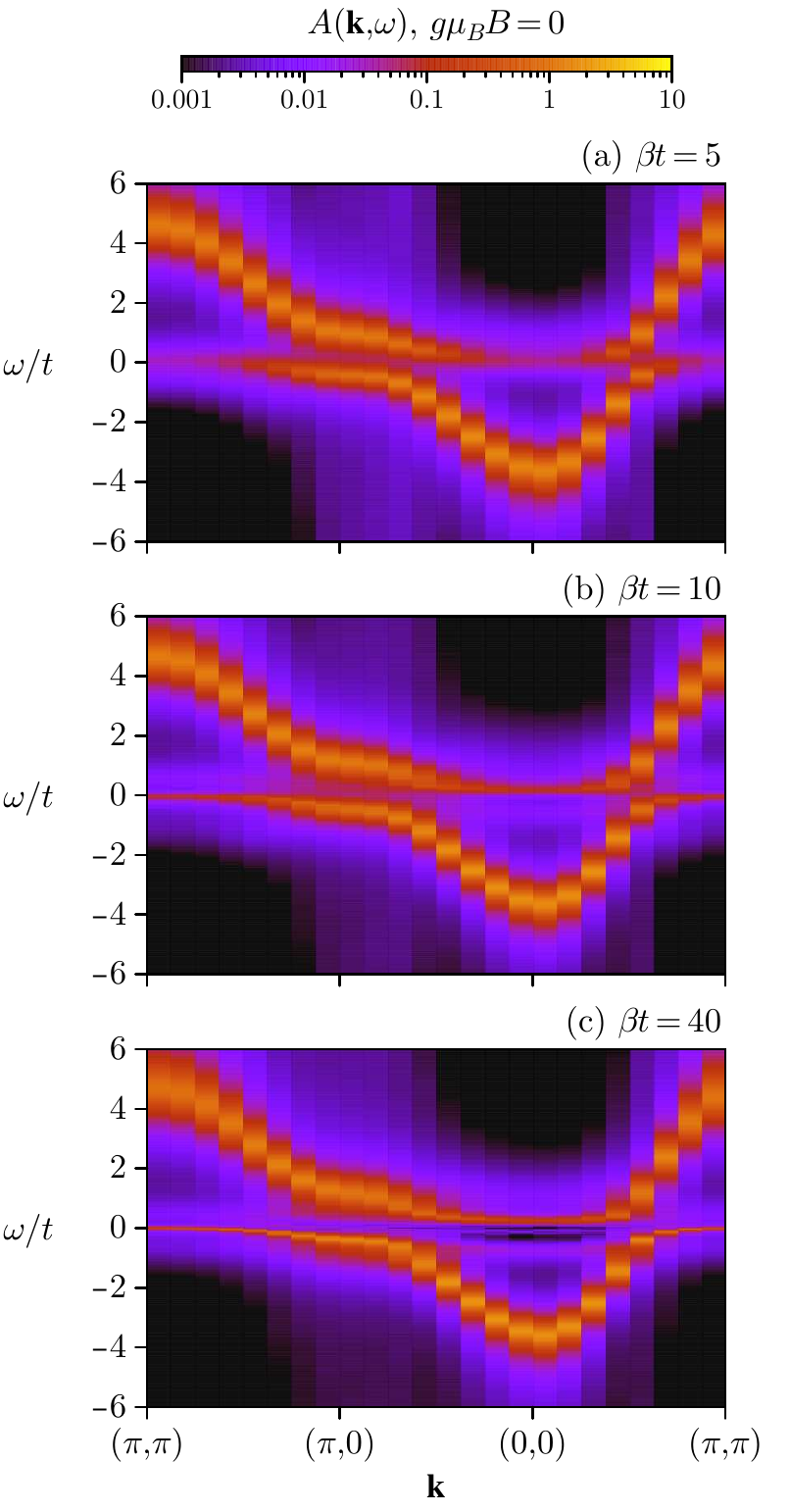}
\end{center}
\caption{\label{FIG:Aom_T}
The $c$-electron spectral function $A(\vec{k},\omega)$ in zero field is plotted for 
three progressively lower temperatures. The legend at the top indicates
the false color values.
}
\end{figure}
	
\begin{figure*}
\begin{center}
\includegraphics{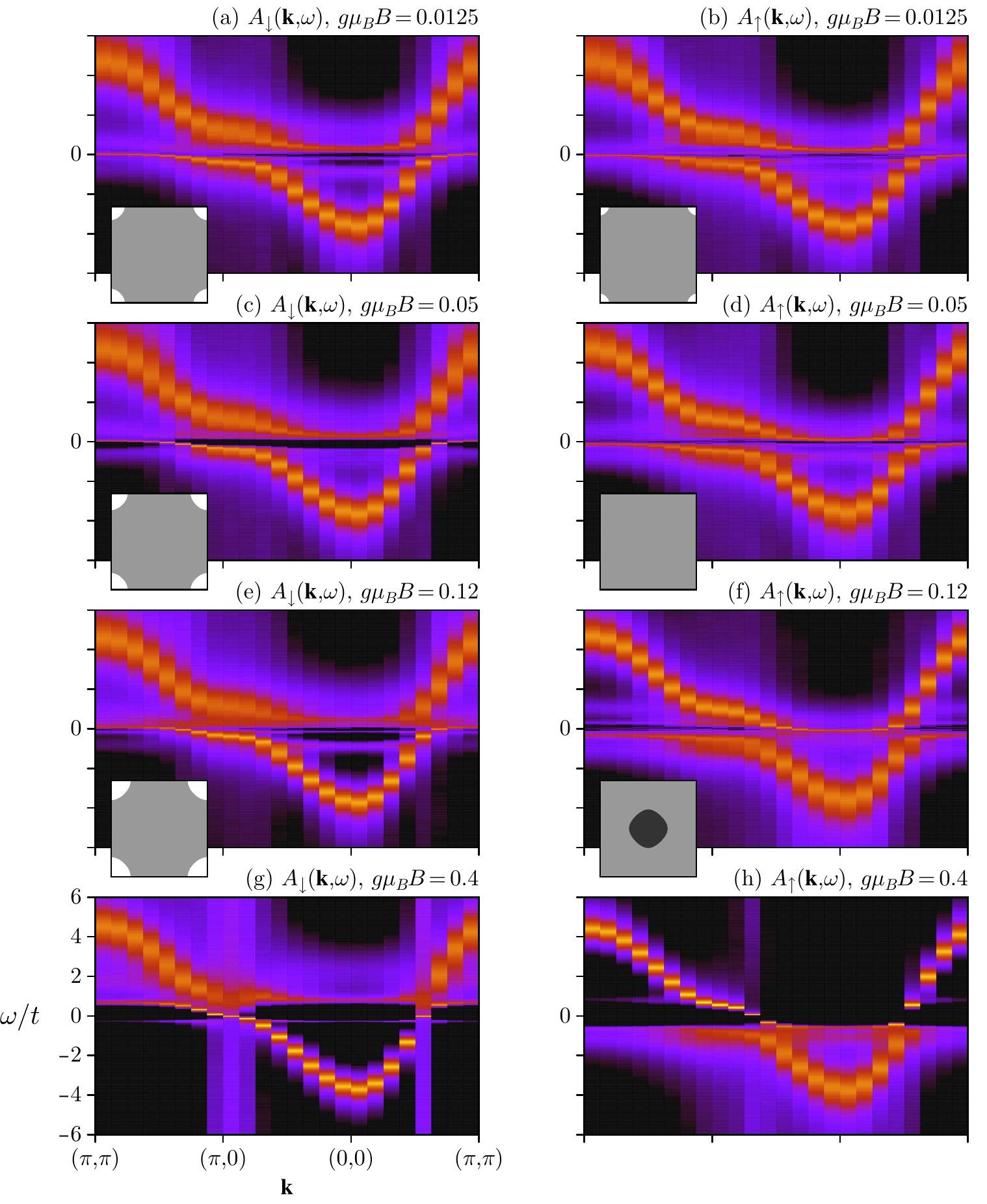}
\end{center}
\caption{\label{FIG:Aom_B}
Low temperature, $\beta t = 40$,  spectral function  in the 
up (left)  and down (right)  spin  sector  as a function of  magnetic field. We use the same scale 
as in Fig.~\ref{FIG:Aom_T}. } 
\end{figure*}

In zero magnetic field and at our lowest temperature, 
$\beta t  = 40$ [see Fig.~\ref{FIG:Aom_T}(c)],
the single particle spectral  function follows the hybridized band picture  
of the  mean-field calculation (cf.\ Fig.~\ref{FIG:bands}), reflecting the coherent heavy fermion 
metallic state with  Luttinger volume including both conduction and localized 
electrons and large effective mass. 
Introducing a magnetic field  leads to  competing effects. One possibile outcome is that Kondo 
screening is completely suppressed---thus triggering the breakdown of the heavy-quasiparticle, 
the  heavy fermion coherent state, and the associated large Fermi surface.  
Another is that partial Kondo screening persists and that the heavy fermion 
metallic state survives in some form.  Fig.~\ref{FIG:Aom_B}  supports the latter.  In particular, in 
region I [Fig.~\ref{FIG:Aom_B}(a),(b)] the two-fold spin degeneracy  of the spectral function 
is lifted and the spin down (up) band shifts  to higher (lower)  energies,  thereby  producing 
hole-like Fermi surfaces of different size centered around $\vec{k} = (\pi,\pi)$. 
[The fact that the spin down (up) band shifts up (down) in energy and yet $m_c < 0$
is a consequence of some subtle rearrangement of spectral weight.]
The onset of region II [Fig.~\ref{FIG:Aom_B}(c),(d)] is marked by the vanishing of the 
spin up Fermi surface: the chemical potential lies within the hybridization gap, and the lower
spin up band is completely filled. This sudden 
change in the Fermi surface topology is the origin of the cusp-like feature separating 
regions I and II.  Note that the  magnetization data of Fig.~\ref{FIG:magnetization}
supports the sharpening of the cusp as the temperature is lowered.  
Region III [Fig.~\ref{FIG:Aom_B}(e),(f)]
is again characterized by a change in the topology of the Fermi surface. 
Here the spin up valence band drops below the chemical potential, thereby forming an 
electron-like spin down Fermi surface centred around the $\vec{k} = (0,0) $ point. 
Across regions I, II, and III, the spin down Fermi surface evolves \emph{continuously} 
with a growing hole-like Fermi surface centered around $\vec{k} = (\pi,\pi)$. 
A sketch of the spin-resolved density  of states 
in the three regions is  given in Fig.~\ref{FIG:magnetization}.
At very high magnetic fields [Fig.~\ref{FIG:Aom_B}(g),(h)] the spectral function progressively tends
towards that of a Zeeman split cosine band. 

The onset of the plateau-like feature in the magnetization curve  as a function of temperature 
can be used as measure of the coherence temperature. At $J/t=1.6$ and $n_c = 0.85$,
Fig.~\ref{FIG:magnetization}  provides a rough estimate of this scale: 
$T_{\text{coh}}/t \simeq 1/30 $. Given $T_{\text{coh}}$, it is interesting to analyze the 
temperature dependence of the single particle spectral function (see Fig.~\ref{FIG:Aom_T}). 
Already at temperatures $T/t = 1/5$ and $T/t=1/10$ above $T_K/t  \simeq 0.09$, features of the 
hybridized bands are apparent. At those temperatures the magnetization  curve is featureless since 
the coherence or hybridization gap is not formed.  At $ T/t = 1/40 < T_{\text{coh}}$ (Fig.~\ref{FIG:Aom_T}c)  the coherence gap is well formed and the plateau feature in the magnetization  curve is 
apparent. 

\begin{figure}
\begin{center}
\includegraphics{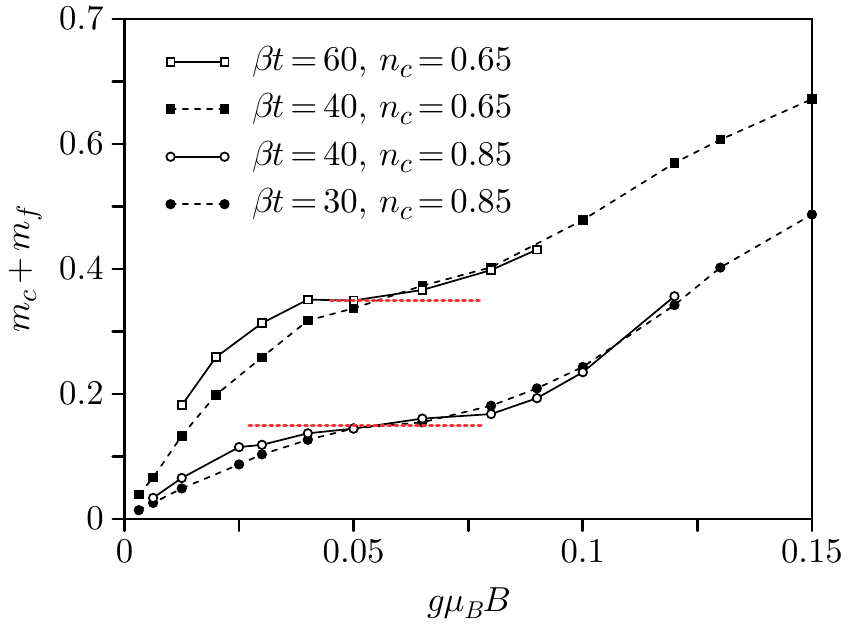} 
\end{center}
\caption{\label{FIG:plateaux} 
Total magnetization, $M = m_c + m_f$,  as function of temperature 
and band filling.   The plateaux at $x=1-n_c$,
predicted by the mean field theory, are indicated with horizontal dashed (red) lines and correspond to 
the vertical dashed (red) lines in the lower panel of Fig.~\ref{FIG:scales}.
}
\end{figure}

As argued in Sec~\ref{SEC:mean-field}, the height of the magnetization plateau, 
$M = 1 - n_c$, follows from  the Luttinger sum rule. Fig.~\ref{FIG:plateaux} 
plots the total magnetization $M = m_c + m_f $ at two different conduction band 
fillings, $n_c = 0.85$ and $n_c = 0.65$. As the temperature decreases, a plateau-like 
feature precisely  at $ 1 - n_c$ emerges. For comparison, we have plotted the 
mean-field prediction as obtained from Fig.~\ref{FIG:scales}.
Note that the initial slope in the magnetization 
curve is related to the effective mass, which is inversely proportional to the coherence temperature. 
Hence, at low band fillings, lower temperature values are required to reveal the plateau feature.  

\section{Conclusions}

The interaction of conducting electrons with local impurities often gives rise to complicated
nonlinearities in magnetic response (often described as metamagnetism).
Below the coherence temperature, we find that  the magnetization profile of 
the Kondo lattice system shows a plateau, whose onset is linked to the vanishing of the 
Fermi surface in one spin sector. The relevant energy scale corresponds to the headroom 
above the chemical potential in the lower hybridized band. As a consequence of the 
Luttinger sum rule, the total magnetization is locked to the value $M = 1 - n_c$
throughout the plateau. The plateau survives up to magnetic fields at which the 
quasiparticles begin to break apart. At high fields, the system is once again characterized 
by Fermi surfaces in both spin sectors, which smoothly evolve with increasing field 
back to those expected for the Zeeman-split bare conduction band. 

We have reached these conclusions on the basis of dynamical and static mean field calculations,
both of which are consistent with the scenario we have outlined.
There is one important disagreement between the two theories.
At large fields, the DMFT shows that once the chemical has traversed the gap, it enters 
the upper hybridized band, and the system again becomes a heavy metal with
two Fermi surfaces. The behavior of the static mean field equations after the reentry
into the upper band is pathological. Energy considerations suggest a first-order
collapse back to the normal state. There is no evidence of this within DMFT,
and we expect that phase fluctuations of the hybridization field are responsible
for washing out this false transition.

In closing, this work establishes that the metamagnetic behavior of Kondo systems 
is intimately related to coherence. We have shown that the magnetization profile of 
a Kondo system is a useful probe of the onset of coherence at low temperatures. 
The location of the incipient plateau reveals the important energy scales;
its height is a direct measure of the Luttinger sum rule.

\acknowledgments  The  simulations were carried out on the IBM p690 and IBM Blue Gene/L
at the John von Neumann Institute for Computing, J\"ulich. We would like to thank this institution 
for generous allocation of CPU time. This work was financially supported by the Alexander von Humboldt Foundation and by the DFG under grant number AS 120/4-2.

\appendix

\section{\label{SEC:DOS}Bare Density of States}

Numerical evaluation of the bare DOS is readily accomplished using
a power series expansion about the Van Hove singularity at $\omega=0$:
\begin{widetext}
 \begin{equation}
\rho_0(\omega) = \frac{1}{2\pi^2t}\biggl[ \log\frac{16t}{|\omega|}
+\biggl(\log\frac{16t}{|\omega|}-1\biggr)\biggl(\frac{\omega}{8t}\biggr)^2
+\frac{9}{4}\biggl(\log\frac{16t}{|\omega|}-\frac{7}{6}\biggr)\biggl(\frac{\omega}{8t}\biggr)^4
+\frac{25}{4}\biggl(\log\frac{16t}{|\omega|}-\frac{37}{30}\biggr)\biggl(\frac{\omega}{8t}\biggr)^6
+\cdots \biggr]
 \end{equation}
 \end{widetext}
Truncating the expansion at eighth order in $\omega/t$ results in a
relative error of at most $10^{-7}$ in the region $-t < \omega < t$
and $10^{-2}$ in the region $-4t < \omega < 4t$.

\end{document}